\documentclass[iop]{emulateapj}
\usepackage{graphicx}

\newcommand{\be}{\begin{eqnarray}}
\newcommand{\ee}{\end{eqnarray}}
\newcommand{\lp}{\left(}
\newcommand{\rp}{\right)}

\newcommand{\E}[1]{\times10^{#1}}

\newcommand{\msol}{ \ M_\odot }
\newcommand{\commentOut}[1]{}
\newcommand{\bi}{\begin{itemize}}
\newcommand{\ei}{\end{itemize}}
\newcommand{\cgsd}{ {\rm \ g \ cm^{-3}}}
\newcommand{\cgsv}{ {\rm \ cm \ s^{-1}}}

\shorttitle{HELIUM DETONATIONS ON WHITE DWARFS}
\shortauthors{SHEN \& MOORE}
\slugcomment{Accepted for publication in The Astrophysical Journal}

\begin{document}


\title{The Initiation and Propagation of Helium Detonations in White Dwarf Envelopes}

\author{Ken J. Shen\altaffilmark{1,2}}
\altaffiltext{1}{Department of Astronomy and Theoretical Astrophysics Center, University of California, Berkeley, CA 94720, USA.}
\altaffiltext{2}{Einstein Fellow.}

\author{Kevin Moore\altaffilmark{3}}
\altaffiltext{3}{Department of Applied Mathematics and Statistics, University of California, Santa Cruz, CA 95064, USA.}


\begin{abstract}

Detonations in helium-rich envelopes surrounding white dwarfs have garnered attention as triggers of faint thermonuclear ``.Ia'' supernovae and double detonation Type Ia supernovae.  However, recent studies have found that the minimum size of a hotspot that can lead to a helium detonation is comparable to, or even larger than, the white dwarf's pressure scale height, casting doubt on the successful ignition of helium detonations in these systems.  In this paper, we examine the previously neglected effects of C/O pollution and a full nuclear reaction network, and we consider hotspots with spatially constant pressure in addition to constant density hotspots.  We find that the inclusion of these effects significantly decreases the minimum hotspot size for helium-rich detonation ignition, making detonations far more plausible during turbulent shell convection or during double white dwarf mergers.  The increase in burning rate also decreases the minimum shell mass in which a helium detonation can successfully propagate and alters the composition of the shell's burning products.  The ashes of these low mass shells consist primarily of silicon, calcium, and unburned helium and metals and may explain the high-velocity spectral features observed in most Type Ia supernovae.

\end{abstract}



\section{Introduction}

He-rich accretion onto a white dwarf (WD) can, for the right range of parameters, lead to the formation of a supersonic He-burning detonation in the accreted shell.  While it was first suggested over 30 years ago that the He detonation can trigger a core detonation and subsequent Type Ia supernova (SN Ia; e.g., \citealt{woos80a,nomo82b}), this ``double detonation'' scenario fell mostly out of favor for years.  However, interest in the scenario has been rekindled in the past decade \citep{fhr07,fink10,sim10,sim12,krom10,ruit11,ruit14a,moll13a,dan14a}, in part due to the possibility that the progenitor system is a double WD binary in which the He detonation occurs during a convective shell burning phase \citep{bild07,sb09b} or during the initial stages of a merger \citep{guil10,rask12,pakm13a}.  The double detonation scenario is able to match observed SN Ia rates \citep{maoz11,ruit11,ruit14a} and avoid unobserved effects that a large non-degenerate binary companion would impart on the SN Ia's light curve and spectra \citep{kase10,bloo12,sp12}.  While recent observations of circumstellar material in $\sim 20\%$ of SNe Ia \citep{pata07,ster11,ster14a,magu13} have been used as evidence for a single degenerate scenario \citep{pata11a,mb12}, these signatures can also be produced during double WD binary evolution \citep{rk13a,shen13a}.

In recent years, it has been suggested that He shell detonations are possible in small shells that might not trigger core detonations and Type Ia supernovae, especially if the donor is a low mass He WD for which the accretor's convective He-burning shell is $ \le 0.1 \msol$.  While subsequent work finds that even detonations in these small He shells will trigger detonations in C/O cores $\gtrsim 0.8 \msol$, O/Ne and lower mass C/O cores may remain intact \citep{shen14a}.  The resulting explosion of only the He shell would be a faint and rapidly evolving ``.Ia'' supernova \citep{bild07,sb09b,shen10,wald11}.  These ``.Ia'' supernovae have been suggested as possible explanations for the newly discovered classes of SNe Iax (see \citealt{fole13a} for an overview), Ca-rich / O-poor transients \citep{pere10,kasl12}, and rapidly fading Type I supernovae (e.g., \citealt{pere11b,drou14a}).

The possibility of double detonations and ``.Ia'' supernovae is predicated on the successful ignition and propagation of the He detonation.  Initial estimates assumed ignition occurs when the local He-burning timescale becomes shorter than the local dynamical timescale, $t_{\rm dynamical} = H/c_s$, where $H$ is the pressure scale height, and $c_s$ is the sound speed \citep{bild07,sb09b,guil10,dan14a}.  However, this is equivalent to assuming that material encompassing the entire scale height is involved in the initiation of the He detonation, so that the timescale for the growing overpressure to expand is the time for sound waves to traverse the scale height.  In actuality, the He detonation is triggered in a very small region within the white dwarf's envelope, so that the actual dynamical timescale of interest is the much shorter sound crossing time of this subregion.

As a result, more recent studies such as \cite{holc13a} quantifying the necessary conditions for He ignition have found that triggering a He detonation is instead prohibitively difficult.  They found that detonations arising from perturbed hotspots require hotspots comparable in size to the white dwarf's scale height.  Given the improbability of generating a stochastic fluctuation as large as this, these studies implied that He detonations do not occur on white dwarfs.

However, because their work was an initial exploration, \cite{holc13a} assumed a pure He composition for their calculations.  In reality, the He layer will be polluted with a non-negligible fraction of C/O.  In the case of He detonations triggered during a double WD merger, such pollution occurs due to dynamical mixing between the direct impact accretion stream and the accretor's core.  For convectively ignited He detonations, the convective fluid motions prior to the detonation may shear across the composition discontinuity between the He shell and C/O core and dredge up core material, in close analogy to metal-enriched classical nova ejecta \citep{gehr98}.  Furthermore, a C/O mass fraction of $5- 10\%$ is generated by the previous He-burning phase prior to the onset of dynamical burning \citep{sb09b}.  In addition to $^{12}$C and $^{16}$O, the accreted material will also contain a significant amount of $^{14}$N, since this isotope is the slowest point of the CNO cycle and will be present in the He-rich material that has undergone prior CNO H-burning.

As we demonstrate in Section \ref{sec:motivation}, these pollutants lead to a significant boost in He-burning rates when a full nuclear reaction network is utilized, allowing for nuclear reactions that bypass the relatively slow triple-$\alpha$ process.  Using the range of temperatures and densities motivated in Section \ref{sec:expectations}, we reexamine the issue of He detonation ignition in more detail in Section \ref{sec:hotspotsizecalculation}.  In addition to constant density hotspots, we also consider constant pressure hotspots that are more appropriate to the subsonic regions where these detonations develop.  We show in Section \ref{sec:hotspotsizeresults} that these amendments to the previous calculations drastically decrease the minimum size of hotspots that give rise to He detonations, making their realization in WD envelopes far more likely.  In Section \ref{sec:shellpropagation}, we calculate the propagation of these detonations within He shells, allowing for post-shock radial expansion and including the previously mentioned nucleosynthetic effects.  We find that propagating He detonations in the smallest allowable shells yield $^{28}$Si and $^{40}$Ca as their main burning products, which may explain the high-velocity features seen in many SNe Ia.


\section{Polluted helium-burning with a large nuclear reaction network}
\label{sec:motivation}

The majority of initial work on He detonations has assumed initial compositions of pure He.  However, as previously noted (e.g., \citealt{wbs06,sb09b,wk11}), $\alpha$-captures onto $^{12}$C and $^{16}$O seed nuclei are far more rapid at the relevant temperatures than the triple-$\alpha$ reaction, especially if the nuclear reaction network used for the calculation includes the proton-catalyzed $\alpha$-capture $^{12} {\rm C}(p,\gamma)^{13}{\rm N}(\alpha,p)^{16}{\rm O}$.  As mentioned previously, small amounts of $^{12}$C and $^{16}$O are expected in the He-rich envelopes at the time of detonation initiation.  While the initial proton abundance is likely very small, protons are also released in $(\alpha,p)$ reactions involving the accreted $^{14}$N as well as $\alpha$-chain nuclei such as $^{20}$Ne, $^{24}$Mg, and $^{28}$Si.

\begin{figure}
	\plotone{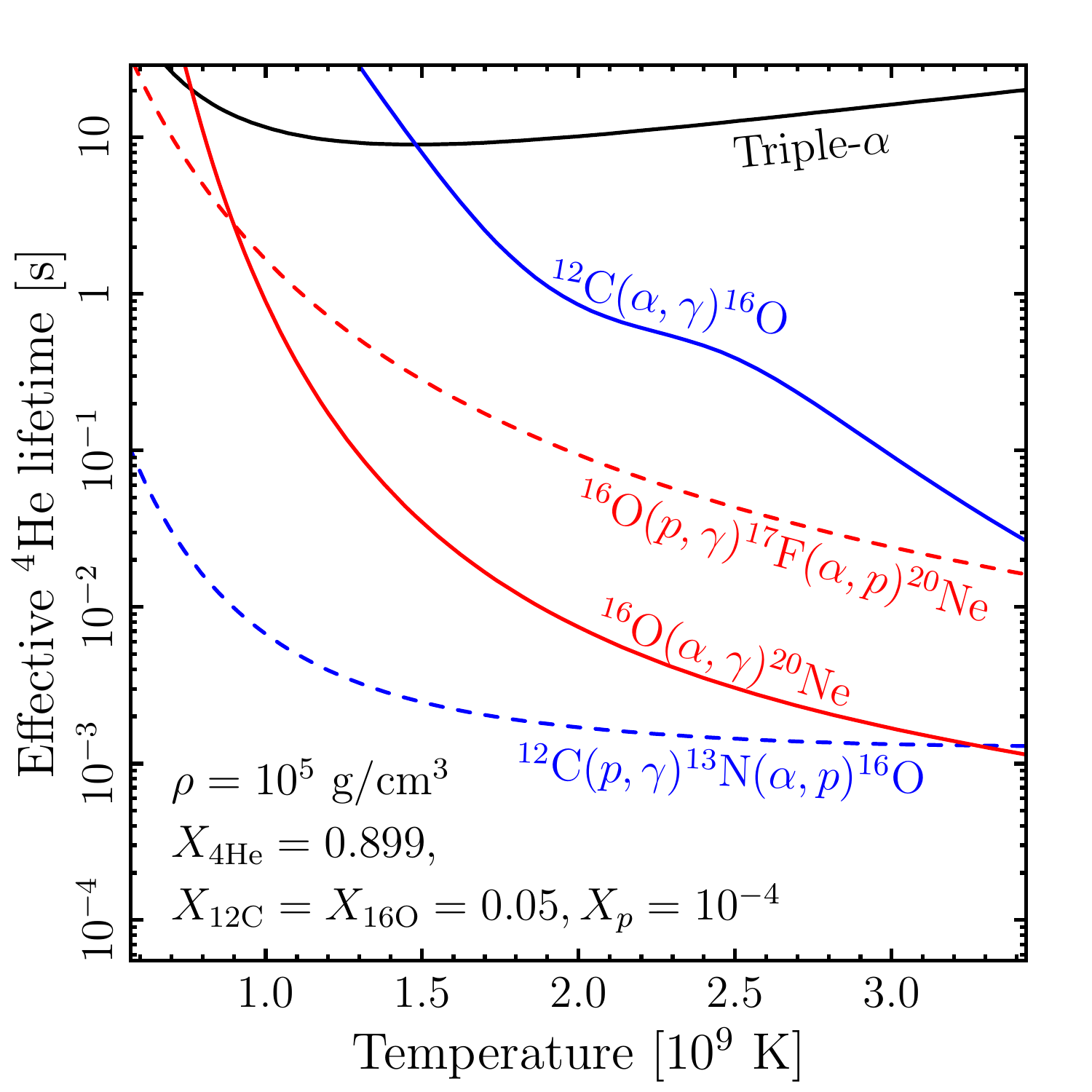}
	\caption{Lifetimes of He nuclei due to various nuclear reactions, as labelled, vs.\ temperature.  The composition and density are given in the figure.}
	\label{fig:acapture}
\end{figure}

Figure \ref{fig:acapture} shows He nuclei lifetimes vs.\ temperature for several direct and indirect $\alpha$-capture reactions at a density of $\rho=10^5 \cgsd$, where the lifetime is defined as $\left| d \ln X_{\rm 4He}/dt \right|^{-1}$ due to reactions with the appropriate target isotope.  The mass fractions of He, $^{12}$C, $^{16}$O, and protons are $X_{\rm 4He}=0.899$, $X_{\rm 12C}=X_{\rm 16O}=0.05$ and $X_p=10^{-4}$.  The lifetimes for the proton-catalyzed reactions are calculated by assuming the intermediate nuclei ($^{13}$N and $^{17}$F) are in reaction rate equilibrium.  It is clear that pure $\alpha$-captures onto $^{12}$C and $^{16}$O dominate over the triple-$\alpha$ process for temperatures $\gtrsim 10^9$ K.  Even more striking is the reduction in the lifetime of He nuclei by four orders of magnitude due to proton-catalyzed $\alpha$-captures onto $^{12}$C seed nuclei.

\begin{figure}
	\plotone{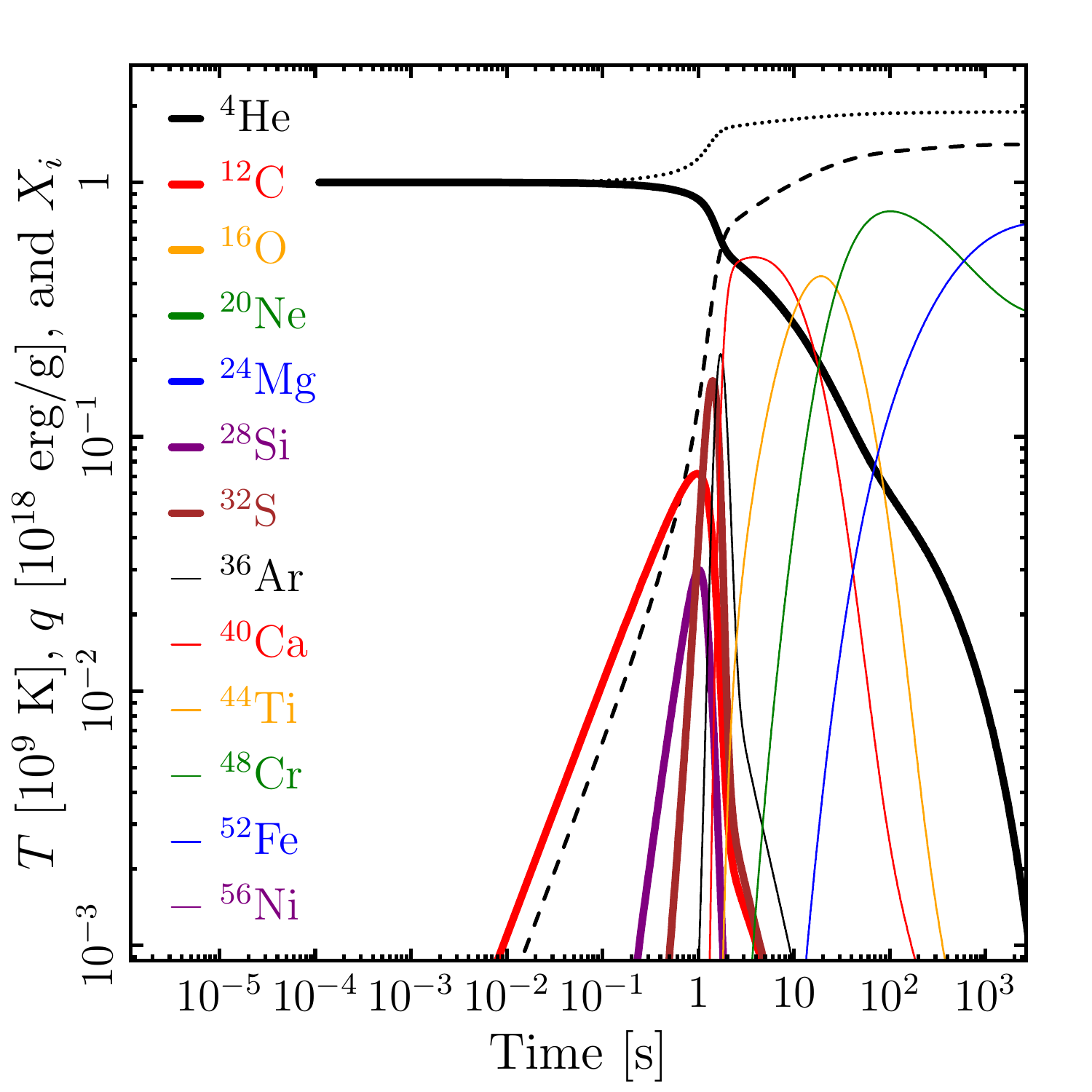}
	\caption{Mass fractions (solid lines), temperature (dotted line), and nuclear energy released (dashed line) vs.\ time.  The initial composition is pure He, and the aprox13 nuclear reaction network is utilized.  Apart from He, which is the black solid line that begins at $X_{\rm 4He}=1$, the solid lines, from left to right in order of their their first appearance in time above a value of $10^{-3}$,  represent $^{12}$C, $^{28}$Si, $^{32}$S, $^{36}$Ar, $^{40}$Ca, $^{44}$Ti, $^{48}$Cr, and $^{52}$Fe.  The mass fractions of $^{16}$O, $^{20}$Ne, $^{24}$Mg, and $^{56}$Ni do not attain values $ \ge 10^{-3}$.}
	\label{fig:xvst_pureHe_a13}
\end{figure}

\begin{figure}
	\plotone{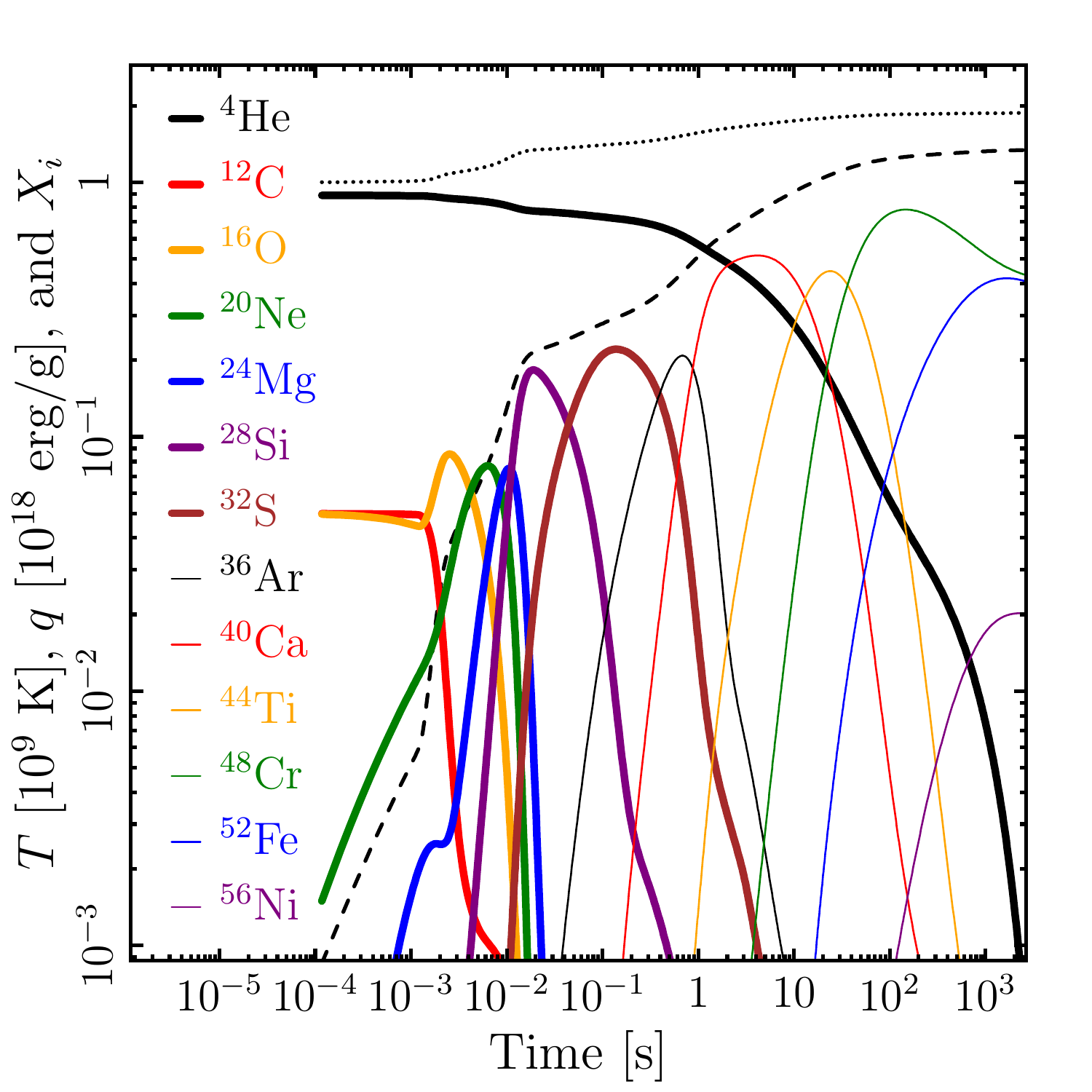}
	\caption{Same as Figure \ref{fig:xvst_pureHe_a13}, but with a 206-isotope nuclear network and an initial composition of $X_{\rm 4He}=0.891$, $X_{\rm 12C}=X_{\rm 16O}=0.05$, and $X_{\rm 14N}=0.009$.  The black solid line beginning near a value of 1 represents the mass fraction of He, and the two solid lines beginning at $0.05$ are $^{12}$C, which is depleted first, and $^{16}$O.  The other solid lines represent the remaining $\alpha$-chain isotopes in ascending mass order from their first appearance in time above a value of $10^{-3}$.  The electron fraction remains essentially unchanged during the calculation from its initial value of $0.5$.  Note that non$-\alpha$-chain isotopes are not displayed, even though some attain values $>10^{-3}$ during the course of the calculation.}
	\label{fig:xvst_HeCNO_206}
\end{figure}

The inclusion of this reaction and the relevant isotopes and reactions that enable it results in a dramatic boost in the He-burning rate.  Figures \ref{fig:xvst_pureHe_a13} and \ref{fig:xvst_HeCNO_206} show mass fractions (thick and thin solid lines; see figures for the association of lines to isotopes), temperature (dotted lines), and nuclear energy release (dashed lines), $q$, as a function of time for two one-zone nuclear burning calculations.  In these one-zone burns, the density is held constant at a value of $10^5 \cgsd$, and the composition and temperature are allowed to vary.  The temperature thus obeys the evolution equation
\be
	\frac{dT}{dt} = \frac{ \epsilon_{\rm nuc} - \epsilon_{\nu}}{c_V} ,
\ee
where  the nuclear energy generation rate, $\epsilon_{\rm nuc} $, the neutrino cooling rate, $\epsilon_{\nu}$, and the specific heat at constant volume, $c_V$, all depend on the changing composition and temperature.

The initial composition in Figure \ref{fig:xvst_pureHe_a13} is pure He, the initial composition in Figure \ref{fig:xvst_HeCNO_206} is $X_{\rm 4He}=0.891$, $X_{\rm 12C}=X_{\rm 16O}=0.05$, and $X_{\rm 14N}=0.009$, and the initial temperature for both figures is $10^9$ K.  The calculations here and throughout the rest of this paper utilize modules included with the MESA\footnote{http://mesa.sourceforge.net, version 5596} stellar evolution code \citep{paxt11,paxt13} for the implicit Rosenbrock integrator \citep{hair96a}, nuclear reaction rates \citep{cybu10a}, neutrino cooling rates \citep{itoh96}, and the equation of state \citep{ts00b}.

The nuclear burning network in Figure \ref{fig:xvst_pureHe_a13} is an often-used 13-isotope network, ``aprox13'' \citep{timm99}, which includes the 13 $\alpha$-chain isotopes, forward and reverse $\alpha$-captures, and $(\alpha,p)(p,\gamma)$ forward and reverse reactions for isotopes $^{24}$Mg and heavier.  In order to allow for $(p,\gamma)(\alpha,p)$ reactions at lower mass numbers, and any other possibly important reactions, the calculation shown in Figure \ref{fig:xvst_HeCNO_206} utilizes a 206-isotope nuclear network that tracks the abundances of neutrons, $^{1-3}$H, $^{3-4}$He, $^{6-7}$Li, $^{7-10}$Be, $^{8-11}$B, $^{11-14}$C, $^{13-15}$N, $^{14-19}$O, $^{17-19}$F, $^{18-23}$Ne, $^{21-24}$Na, $^{22-27}$Mg, $^{25-28}$Al, $^{27-32}$Si, $^{29-34}$P, $^{31-37}$S, $^{33-38}$Cl, $^{36-41}$Ar, $^{37-42}$K, $^{40-49}$Ca, $^{41-50}$Sc, $^{44-51}$Ti, $^{45-52}$V, $^{48-55}$Cr, $^{51-57}$Mn, $^{52-61}$Fe, $^{55-62}$Co, $^{56-65}$Ni, $^{57-66}$Cu, $^{60-69}$Zn, $^{61-70}$Ga, and $^{64-71}$Ge, and all of their interlinking nuclear reactions.

It is evident from a comparison of the two figures that the addition of a small amount of CNO isotopes and the use of a large reaction network vastly shortens the time to release a significant amount of energy.  The polluted full network calculation reaches $q = 2\E{17}$ erg g$^{-1}$ more than 100 times faster than the pure He aprox13 case.

\begin{figure}
	\plotone{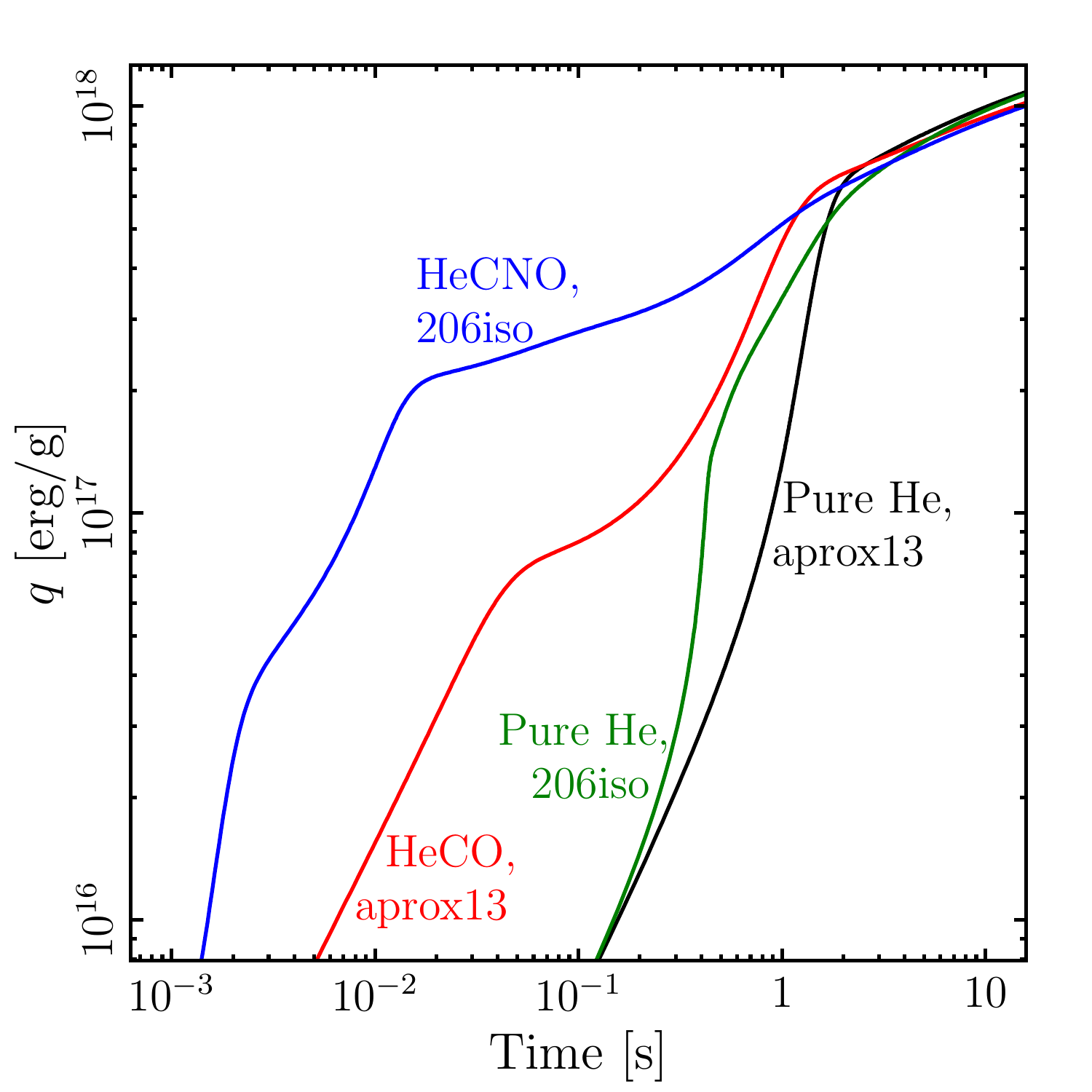}
	\caption{Nuclear energy release, $q$, vs.\ time for four combinations of initial compositions and nuclear reaction networks, aprox13 and a 206-isotope network.  The initial compositions are pure He, ``HeCO'' ($X_{\rm 4He} = 0.9$ and $X_{\rm 12C}=X_{\rm 16O}=0.05$), and ``HeCNO'' ($X_{\rm 4He} = 0.891$, $X_{\rm 12C}=X_{\rm 16O}=0.05$, and $X_{\rm 14N}=0.009$).  All four calculations begin at a temperature of $10^9$ K and have a constant density of $10^5 \cgsd$.}
	\label{fig:qvst}
\end{figure}

This speeding up of the energy release is demonstrated in Figure \ref{fig:qvst}, which compares the results of four combinations of nuclear reaction networks and initial compositions.  Two of the lines (``Pure He, aprox13'' and ``HeCNO, 206iso'') are as in Figures \ref{fig:xvst_pureHe_a13} and \ref{fig:xvst_HeCNO_206}, respectively.  The ``Pure He, 206iso'' calculation begins with an initial composition of pure He and utilizes the same 206-isotope network as in Figure \ref{fig:xvst_HeCNO_206}.  The ``HeCO, aprox13'' calculation uses the aprox13 network and a polluted initial composition that mimics that of Figure \ref{fig:xvst_HeCNO_206}, but since $^{14}$N does not exist in the aprox13 network, the initial composition is $X_{\rm 4He}=0.9$ and $X_{\rm 12C}=X_{\rm 16O}=0.05$.  While a significant decrease in the burning timescale is seen in the polluted calculation with aprox13, a complete reaction network is required to gain the full boost.


\section{Expectations for hotspot conditions}
\label{sec:expectations}

Before exploring the effects of these decreased He-burning timescales on the initiation and propagation of detonations, we first estimate the characteristics of hotspots created during double WD mergers and during convective shell burning on WD surfaces to motivate the range of hotspot conditions we consider in Section \ref{sec:hotspotsizecalculation}.


\subsection{Convective hotspots}
\label{sec:convhotspots}

He detonations may arise in convective He-burning shells as they approach the point of inefficient convection, when the eddy turnover timescale becomes comparable to the local burning timescale, and the assumption of a globally isentropic envelope breaks down \citep{shen10}.  We estimate the spectrum of temperature fluctuations in a convective shell by utilizing similar arguments to those employed for the situation of convective C-burning in WD cores (e.g., \citealt{woos07,pan08a,schm10a}).

We assume that the probability density function of turbulent energy dissipation fluctuations in the convective zone follows a log-normal distribution \citep{kolm41a,kolm62a}:
\be
	{\rm pdf} \left[ \ln{ \lp \epsilon_l/ \epsilon_H \rp } \right] = \frac{1}{\sqrt{ 2 \pi \sigma_l^2}} \exp{ \left[ - \frac{ \lp \ln{ \lp \epsilon_l/ \epsilon_H \rp } + \sigma_l^2/2 \rp^2 }{2 \sigma_l^2} \right] } ,
\ee
where $\sigma_l^2 \simeq 0.2 \ln{ \lp H/l \rp}$ \citep{schm10a}, the integral scale height of the convective zone is $H$, and $l$ is the inertial lengthscale of interest.  The mean energy dissipation rate at the largest scale is $\epsilon_H = v_{\rm conv}^3/H$, where the convective velocity at the integral scale height is $v_{\rm conv}$.  Turbulent velocities at inertial lengthscales are $v_l = v_{\rm conv} (l/H)^{1/3}$.  The probability distribution function is normalized such that
\be
	\int_{- \infty}^{ \infty} {\rm pdf}  \left[ \ln{ \lp \epsilon_l/ \epsilon_H \rp } \right]  d \ln ( \epsilon_l / \epsilon_H)=1 .
\ee
The probability of obtaining a fluctuation $\epsilon < \epsilon_{\rm fluc}$ is then
\be
	P(\epsilon < \epsilon_{\rm fluc}) = \frac{1}{2} + \frac{1}{2} {\rm erf} \left[ \frac{  \ln \lp \epsilon_{\rm fluc}/\epsilon_H \rp + \sigma_l^2/2 }{ \sqrt{ 2 \sigma_l^2} } \right] .
\ee
The log-normal probability density function is not a complete description of intermittent turbulence \citep{she94a,pan08a,schm10a}, but we utilize it here for simplicity and defer a more careful analysis to future work.

We wish to know the scale of the maximum dissipation fluctuation expected at a given lengthscale within the convective zone, $\epsilon_{\rm max}(l)$.  We estimate this by calculating the probability that none of the eddies with lengthscale, $l$, has a fluctuation larger than $\epsilon_{\rm max}(l)$, and setting this probability equal to 50\%.  Within the volume of the convective zone, there are 
\be
	N_l = \frac{4 \pi R^2 H}{4 \pi l^3 /3} = \frac{3 R^2 }{H^2} \lp \frac{H}{l} \rp^3
\ee
eddies of size, $l$, where $R$ is the WD's radius.  Furthermore, the turbulent cascade at the lengthscale, $l$, is reset every local eddy turnover timescale, $t_l=l/v_l$.  There are then $N_t$ new instantiations during the global eddy turnover timescale, $t_{\rm global}=H/v_{\rm conv}$, where
\be
	N_t = \frac{t_{\rm global}}{t_l} =  \frac{ H/v_{\rm conv} }{ l/v_l} = \lp \frac{H}{l} \rp^{2/3} .
\ee
Thus, $\epsilon_{\rm max}(l)$ is implicitly given by
\be
	\frac{1}{2} = \lp \frac{1}{2} + \frac{1}{2} {\rm erf} \left[ \frac{  \ln \lp \epsilon_{\rm max}(l)/\epsilon_H \rp + \sigma_l^2/2 }{ \sqrt{ 2 \sigma_l^2} } \right] \rp^{N_t N_l} .
\ee

Since the fluctuation only acts over the local eddy turnover timescale before the distribution of eddies is reset, the increase in internal energy from this maximum fluctuation is given by
\be
	\delta u_{\rm max}(l) &=& \lp \epsilon_{\rm max}(l) - \epsilon_H \rp  t_l \nonumber \\
	&=& \lp \frac{ \epsilon_{\rm max}(l)}{\epsilon_H} -1 \rp v_c^2 \lp \frac{l}{H} \rp^{2/3} .
\ee
The resulting expected maximum fluctuations and internal energy changes vs.\ lengthscale are shown in Table \ref{tab:fluc} for typical values of the convective scale height of $10^8$ cm and a radius of $5\E{8}$ cm (suitable for a $0.05 \msol $ envelope on a $1 \msol$ WD as it approaches inefficient convection; \citealt{sb09b}).  The $\delta u_{\rm max} $ values, which scale directly with $v_{\rm conv}^2$, assume $v_{\rm conv} = 10^8 \cgsv$.  Typical values of the integral convective velocity range from $1-2\E{8} \cgsv $.

\begin{table}
	\begin{center}
	\caption{Maximum fluctuation and change in internal energy vs.\ lengthscale}
	\label{tab:fluc}
	\begin{tabular}{|c|c|c|}
	\hline
	$l/H$ & $\epsilon_{\rm max}(l) / \epsilon_H$ & $\delta u (v_{\rm conv}/10^8 {\rm cm/s})^2$ [erg/g] \\
	\hline
	\hline
	0.003 & 850 & $17\E{16}$\\
	\hline
	0.01 & 230 & $11\E{16}$\\
	\hline
	0.03 & 68 & $6.5\E{16}$\\
	\hline
	0.1 & 18 & $3.7\E{16}$\\
	\hline
	0.3 & 5.4 & $2.0\E{16}$\\
	\hline
	\end{tabular}
	\end{center}
\end{table}

For a He-dominated convective zone at a temperature of $3\E{8}$ K and density of $10^5 \cgsd$, the specific heat at constant pressure is $c_P = 1.5\E{8}$ erg g$^{-1}$ K$^{-1}$.  Thus, the maximum temperature fluctuation expected for a lengthscale range of 1-10\% of the scale height, assuming a convective velocity of $1.5\E{8} \cgsv$, is $0.8-1.9\E{9}$ K, with a large range of uncertainty.  Future comparison to numerical hydrodynamic simulations such as that by \cite{zing13a} will help to validate these estimates.


\subsection{Merger hotspots}

He detonations may also arise during the merger of two WDs.  The lower-mass WD may be a He WD, in which case the presence of He is unsurprising.  However, low mass C/O WDs also possess relatively sizable He layers $\sim 10^{-2} \msol$ \citep{it85}, so the merger of two C/O WDs can yield a He detonation if the accreted He layer is large enough.  This possibility was first raised by \cite{rask12} and \cite{pakm13a}, although they did not perform resolved calculations of the He detonation initiation or propagation.  We will see in Section \ref{sec:shellpropagation} that a He detonation can propagate in a He shell as small as $5\E{-3} \msol$ on the surface of a $1.0 \msol $ WD.

To estimate the properties of hotspots produced in mergers, we utilize the results of \cite{dan14a}, who perform a suite of double WD merger simulations with a smoothed particle hydrodynamics code.  For relevant total merger masses $ \gtrsim 1.2 \msol$ and $\lesssim 1.5 \msol$, they find hotspots with maximum temperatures and densities of $ 0.5-1\E{9}$ K and $ 2\E{4} - 10^6 \cgsd$.  However, it is likely that these values are underestimates due to the relatively low resolution of $4\E{4}$ particles in these simulations.  The energy in an under-resolved hotspot is unphysically averaged out over a large mass; higher resolution allows the energy fluctuation to be put into a smaller mass, leading to a higher peak temperature, as seen in resolution studies of WD collision and merger simulations by \cite{ross09a} and \cite{pakm12a}.  With this in mind, and given the estimates for hotspots in convective zones from Section \ref{sec:convhotspots}, we consider a range of hotspot temperatures $0.5-2\E{9}$ K and densities $ 10^5 - 10^6 \cgsd$ throughout the rest of the paper.


\section{Minimum sizes of detonatable helium-rich hotspots}
\label{sec:hotspotsizecalculation}

The spontaneous ignition of detonations from fuel hotspots has been a well-studied topic (e.g., \citealt{zeld70,lee78a,clav04}).  In this section, we estimate the success of detonation ignition via the Zel'dovich gradient mechanism framework \citep{zeld70}, in which the initiation of detonations occurs by a suitable gradient of induction time within a perturbed region, which is due to spatially varying fuel concentrations or thermodynamic variables.  We consider hotspots with initially uniform composition and varying temperature and density gradients in this work; we defer a study of the effects of a spatially inhomogeneous composition to future research.


\subsection{Description of calculation}

We take the temperature profiles within the hotspots to be linear in radius, so that
\be
T(r) = T_{\rm center} - (T_{\rm center} - T_0) \frac{ r}{l_{\rm hotspot}} ,
\ee
where $T_{\rm center}$ is the temperature at the center of the hotspot, $T_0$ is the temperature of the surrounding unperturbed medium, $r$ is the distance from the hotspot's center, and $l_{\rm hotspot}$ is the size of the perturbed region.  Different parameterizations of the thermal profile will lead to somewhat different results; e.g., \cite{seit09} found that using linear, Gaussian, and exponential profiles changes the minimum detonatable size for C/O mixtures by a factor of a few.  Since we only seek an estimate of the critical hotspot sizes, we defer the exploration of different profiles to future work.  The surrounding temperature is taken to be $10^7$ K for our calculations of detonation initiation.  Calculations were also performed with $T_0=10^8$ K, but the resulting minimum hotspot sizes are only altered by tens of percent, so we limit our results to $T_0=10^7$ K.

We consider two types of density profiles: isochoric, or constant density, and isobaric, or constant pressure.  The first is simply a constant density profile: the temperature is the only spatially varying quantity within the hotspot.  For the isobaric case, the density profile varies in such a way as to keep the pressure spatially constant inside and outside the hotspot.

Previous studies have focused on isochoric hotspots.  For densities and temperatures typical of C-burning calculations \citep{al94b,nw97,rwh07a,seit09}, isochoric hotspots are nearly isobaric because the hotspots are mostly electron degenerate, and thus the distinction is not very meaningful.  However, for the lower density conditions suitable for He-burning, the difference can be quite significant.  An isobaric hotspot with a central temperature $T_{\rm center} = 10^9$ K and $\rho_{\rm center} = 10^5 \cgsd$ and a surrounding temperature of $T_0 = 10^7$ K will have a surrounding density of $\rho_0 = 5.0\E{5} \cgsd$.  This reduces the burning lengthscale (often referred to as the Zel'dovich - von Neumann - D\"{o}ring, or ZND, lengthscale) in the surrounding medium by a large factor, which has important implications for minimum hotspot sizes.  We will return to this point in the following sections.

As a result of the induction time gradient, the center of the hotspot burns first, followed by the surrounding region, and so on.  This yields a burning front with an outward velocity equal to the inverse of the induction time gradient, $v_{\rm burn} = (dt_{\rm induction}/dr)^{-1}$.  For a detonation to develop, this burning front velocity should equal the steady-state detonation velocity (hereafter referred to as the Chapman - Jouguet, or CJ, velocity) at a point where at least a ZND lengthscale's worth of material has been burned \citep{zeld70,hc94}.

Our estimate of the minimum size of a hotspot that can transition to a detonation via the Zel'dovich gradient mechanism proceeds in the following way:

1) Specify the nuclear network, composition, central temperature and density of the hotspot, density profile (isochoric or isobaric), and surrounding temperature.

2) Calculate the surrounding density and the density profile within the hotspot if the hotspot is isobaric, assuming a linear temperature profile.

3) Choose a value for $q$, the energy release of the propagating detonation.

4) Calculate the time to release this $q$ at various points within the hotspot using one-zone burns as in Section \ref{sec:motivation}.  These induction times should be sampled densely enough within the hotspot such that they can be used to reliably calculate the induction time gradient.  The induction times are calculated at a fixed density but changing temperature and composition.  This mimics the formation of a supersonic burning front, for which the fluid elements do not have the necessary time to expand and change their density.  Allowing for a changing composition is crucial for the production of trace isotopes (e.g., protons) that dramatically alter burning rates in large nuclear networks at high temperatures.\footnote{A similar acceleration due to the increasing temperature can be accounted for by the Frank-Kamenetskii factor (e.g., \citealt{khok89}) but an analogous factor accounting for the composition is difficult to define.}

5) Calculate the CJ velocity and the ZND lengthscale in the surrounding unperturbed medium for a detonation that releases an energy equal to $q$.  These are estimated by starting with the post-shock, but pre-burned, material and following its time-dependent evolution at constant density.  Here, the assumption of a constant density is not quite correct, as the density does change in a ZND calculation, but given the desired accuracy of our estimates, we do not consider this evolution.

6) Find the size of the hotspot such that the burning wave velocity equals the CJ velocity at a radius where a ZND lengthscale's worth of material has already burned behind it.

7) Repeat for various values of $q$ until the overall minimum detonatable hotspot size is found for the particular initial conditions specified in step 1.

The radius in step 6 is usually just where $r=l_{\rm ZND}$.  However, because the triple-$\alpha$ reaction rate decreases with increasing temperature $\gtrsim 1.5\E{9}$ K, pure He hotspots with high maximum temperatures will not burn most rapidly near the center.  This is especially true for isobaric hotspots where the density decreases towards the center, further increasing the induction time there.  For these situations, we declare a successful detonation initiation if the burning velocity reaches the CJ value at a radius that is $l_{\rm ZND}$ outside of the point of minimum induction time.


\subsection{Example calculation}
\label{sec:hotspotsizeexample}

\begin{figure}
	\plotone{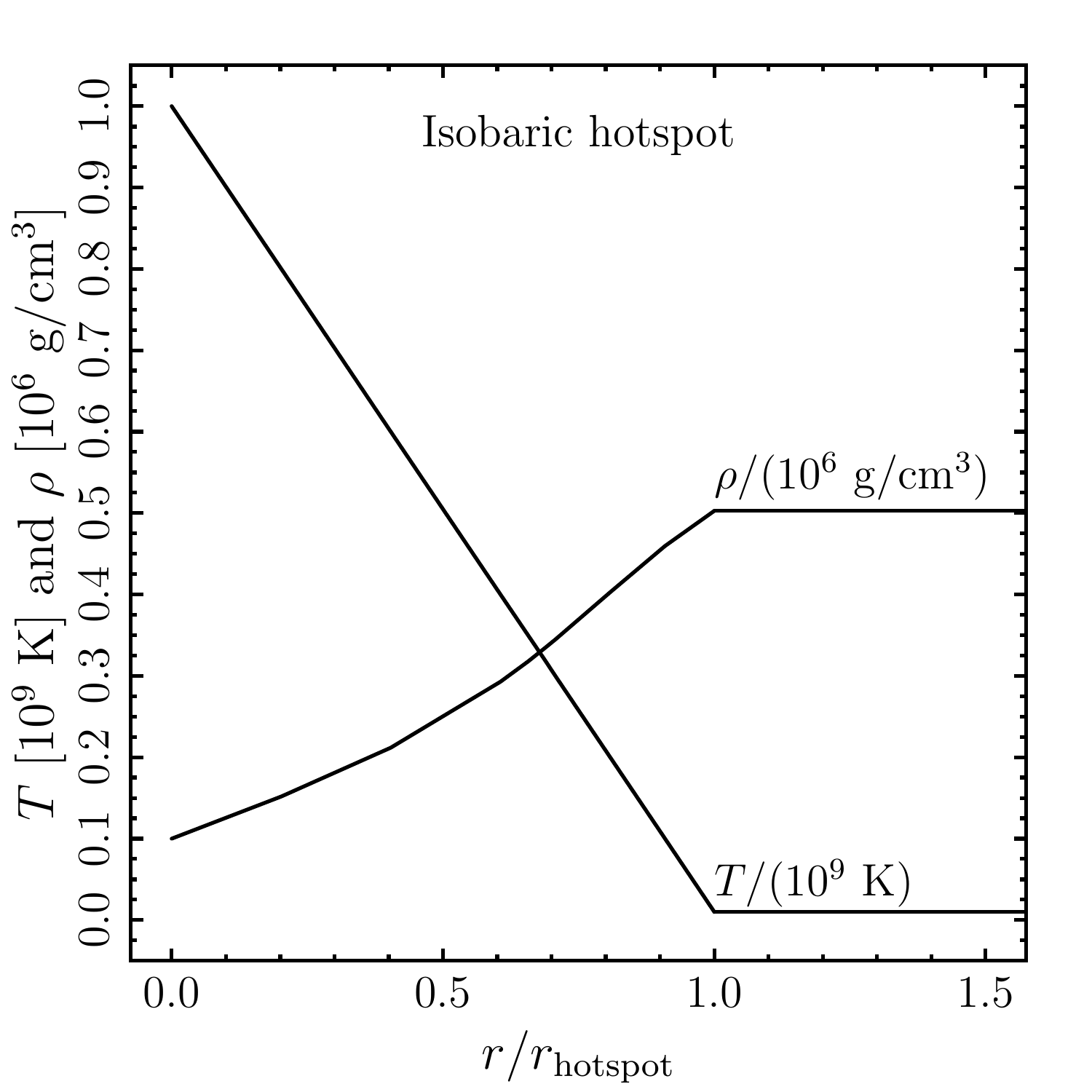}
	\caption{Density and temperature profiles for an isobaric hotspot with a linear temperature profile, $T_{\rm center} = 10^9$ K, $\rho_{\rm center} = 10^5 \cgsd$, and $T_0 = 10^7$ K.  The unperturbed density is $\rho_0=5.0\E{5} \cgsd$.}
	\label{fig:densprof}
\end{figure}

In this section, we demonstrate the process of determining the minimum detonatable hotspot for an isobaric pure He hotspot with a central temperature and density of $T_{\rm center} = 10^9$ K and $\rho_{\rm center} = 10^5 \cgsd$, a surrounding temperature of $T_0 = 10^7$ K, and the aprox13 nuclear network \citep{timm99}.  Figure \ref{fig:densprof} shows the temperature and density profiles within the isobaric hotspot and surrounding unperturbed medium.  The temperature profile is linear between the maximum temperature of $10^9$ K at the center and the unperturbed value of $10^7$ K at the edge of the hotspot.  The density, which is $10^5 \cgsd$ at the center, increases to a value of $5.0\E{5} \cgsv$ at the hotspot's edge to keep the pressure constant.

\begin{figure}
	\plotone{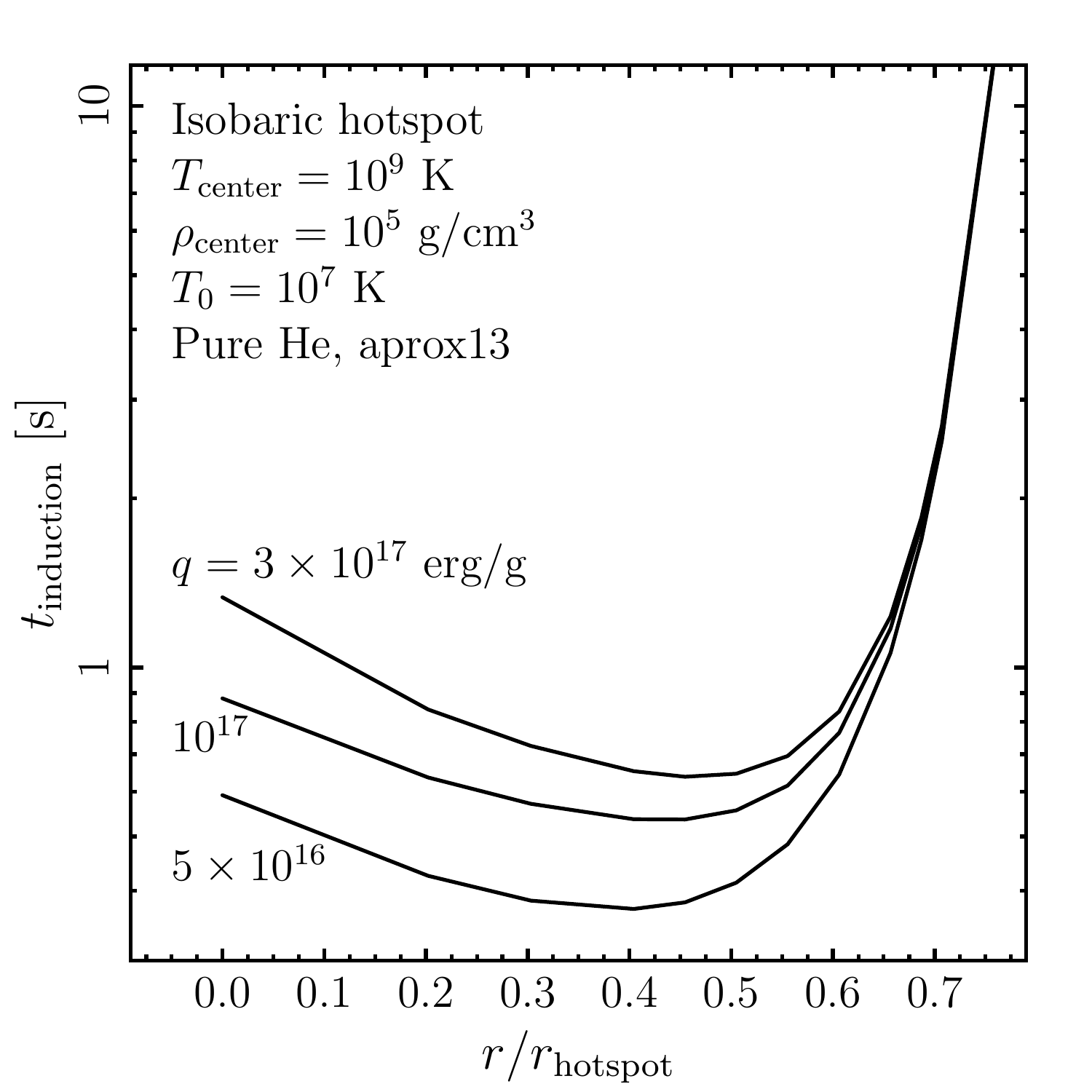}
	\caption{Induction time to release various values of $q$, as labeled, in the hotspot described in Figure \ref{fig:densprof}.  The time evolution of the various points is calculated with the aprox13 network at a constant density and changing temperature and composition.  The initial composition is pure He.}
	\label{fig:tinduction}
\end{figure}

Since the temperature decreases outwards while the density increases, the induction time does not monotonically increase with radius.  This can be seen in Figure \ref{fig:tinduction}, which shows the time to release various values of $q$, as labeled, for the temperature and density profiles in Figure \ref{fig:densprof}.  The minima in induction times are at a radius roughly halfway between the center and the edge of the hotspot.  As mentioned previously, the induction time is found by doing a time-dependent integration of the material at a constant density to simulate the formation of a supersonic wave.

The next step is to calculate the detonation properties in the unperturbed medium.  The velocity of a steady state planar detonation is given by the Chapman-Jouguet value of $v_{\rm CJ} = \sqrt{2 (\gamma^2-1) q}$, where $\gamma$ is assumed to be the value of both adiabatic indices, $\Gamma_1= \left. d\ln P/d\ln \rho \right|_s$ and $\Gamma_3 = 1 + \left. d \ln T / d \ln \rho \right|_s$.  In practice, the values of $\Gamma_1$ and $\Gamma_3$ returned by the Helmholtz equation of state \citep{ts00b} are slightly different, so an average of the two is used for $\gamma$.

For a given $v_{\rm CJ}$, the standard shock jump conditions yield the temperature and density of the shocked but unburned material.  Since the value of $\gamma$ depends on the post-shock temperature and density, we vary the CJ velocity until the derived $\gamma$ is self-consistent.  
The post-shock but pre-burn temperature and density are then used to perform a time-dependent integration at a constant density but changing temperature and composition.  In a proper ZND calculation, the density also changes, but for the sake of simplicity, we hold it constant.

\begin{figure}
	\plotone{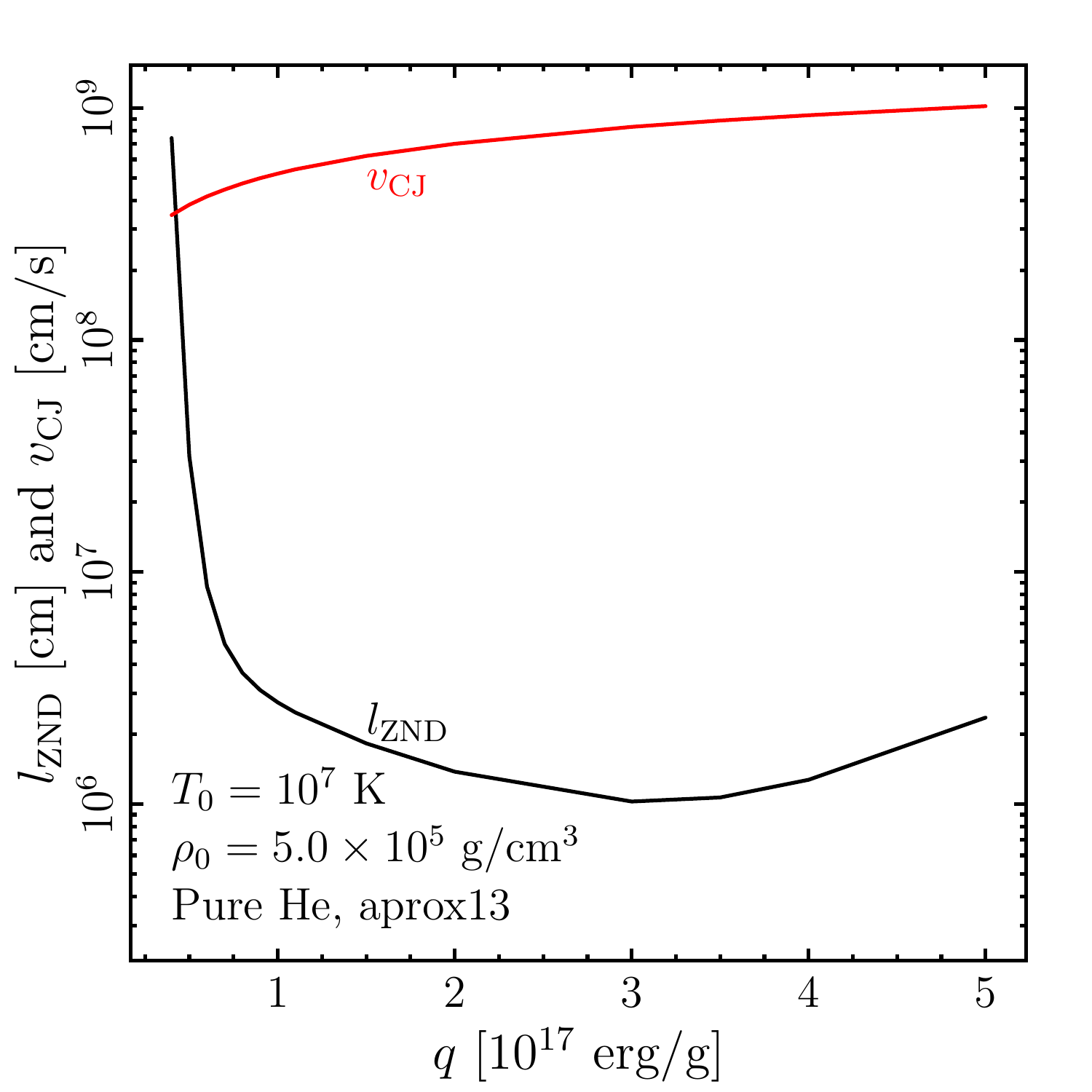}
	\caption{ZND lengthscale and CJ velocity vs.\ $q$ in a medium with $T_0 = 10^7$ K and $\rho_0=5.0\E{5} \cgsd$.  The initial composition is pure He, and the aprox13 nuclear network is utilized.}
	\label{fig:lZNDvsq}
\end{figure}

This calculation yields the induction time of this material to release the assumed $q$.  Multiplying this time by the CJ velocity yields an estimate of $l_{\rm ZND}$, both of which are shown in Figure \ref{fig:lZNDvsq}.  The calculations are now in place for an estimate of the minimum detonatable hotspot size.  For a given $q$, Figure \ref{fig:lZNDvsq} gives the CJ velocity and detonation lengthscale in the unperturbed medium.  The size of the hotspot in Figure \ref{fig:tinduction} is then adjusted until the burning wave velocity is equal to the CJ velocity at a distance that is a detonation lengthscale outside of the point of minimum induction time.  The calculation is then repeated for various values of $q$ until a minimum detonatable hotspot size is found, which, for this fiducial example, is $8.3\E{7}$ cm at $q=2.5\E{17}$ erg g$^{-1}$.


\subsection{Results}
\label{sec:hotspotsizeresults}

\begin{figure}
	\plotone{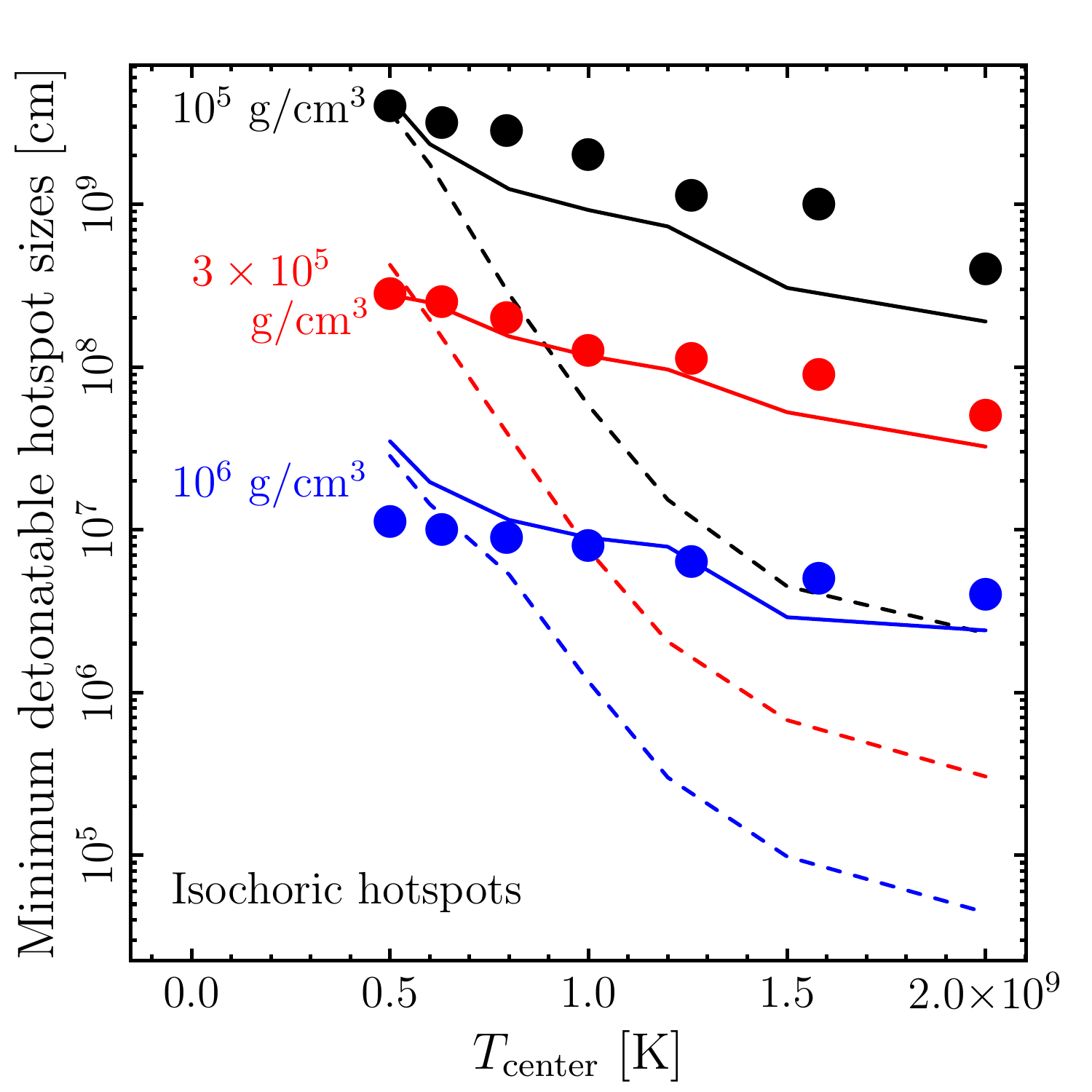}
	\caption{Minimum isochoric hotspot sizes for initiations of detonations vs.\ initial hotspot temperature in He-rich material.  The densities in the hotspots are as labeled.  Bullets show results from the numerical hydrodynamic simulations of \cite{holc13a}.  Solid lines show our results for an initially pure He composition with the aprox13 network for comparison.  Dashed lines show minimum hotspot sizes in an initially He-C-N-O mixture with a 206-isotope network.}
	\label{fig:lcrit_isochoric}
\end{figure}

We now show the results of our Zel'dovich gradient mechanism estimates for minimum detonatable hotspot sizes in He mixtures.  Figure \ref{fig:lcrit_isochoric} shows minimum sizes as a function of the central temperature for constant density hotspots at three labeled densities.  Solid lines show our estimates for pure He media and the aprox13 network.  Bullets show the numerical hydrodynamics results of \cite{holc13a} for comparison.  While the agreement between our results and theirs is not perfect, it is typically within tens of percent, giving us confidence that we are capturing the basic physics of the detonation initiation mechanism.

The dashed lines represent calculations with a 206-isotope network and a polluted mixture with mass fractions $X_{\rm 4He}=0.891$, $X_{\rm 12C}=X_{\rm 16O} = 0.05$, and $X_{\rm 14N}=0.009$.  It is clear that the presence of CNO isotopes and the extended network significantly decrease the minimum detonatable hotspot sizes, particularly at temperatures $ \gtrsim 10^9 $ K.  At the highest temperatures, the minimum hotspot radii decrease to as little as 1\% of the pure He - aprox13 results.  The hotspot size of the fiducial calculation ($T_{\rm center}=10^9$ K and $\rho_{\rm center} = 10^5 \cgsd$) decreases to $6\E{7}$ cm, which is still a fair fraction of a WD's scale height, but is significantly less than the pure He - aprox13 result of $9\E{8}$ cm.

\cite{seit09} also calculated critical hotspot sizes for a very polluted He-rich mixture with $X_{\rm 4He}=0.14$ and $X_{\rm 12C}=X_{\rm 16O}=0.43$ using the aprox13 network.  Their choice of composition was motivated by having one He nucleus for every $^{12}$C nucleus, so that the triple-$\alpha$ reaction can be bypassed entirely.  However, due to the limited network, the $\alpha$-captures onto $^{12}$C and $^{16}$O are still relatively slow.  As a result, their minimum detonatable hotspots are a factor of a few times smaller than the pure He results shown in Figure \ref{fig:lcrit_isochoric} but still larger than our slightly polluted results with a 206-isotope network.

Similarly, \cite{wk11} calculated the minimum sizes of initially pure He hotspots that yield successful detonations while utilizing a large reaction network.  At a density of $10^6 \cgsd$ and temperature of $3.5\E{8}$, they find a minimum detonatable size somewhat larger than $10^7$ cm, which is in rough agreement with an extrapolation of our results in Figure \ref{fig:lcrit_isochoric}.  Note that their choice of central temperatures $2.5 - 3.5\E{8}$ K was motivated by the background temperatures in a convective He shell.  Our significantly hotter range $ \ge 5\E{8}$ K is chosen to match the stochastic temperature fluctuations over the convective mean background.

\begin{figure}
	\plotone{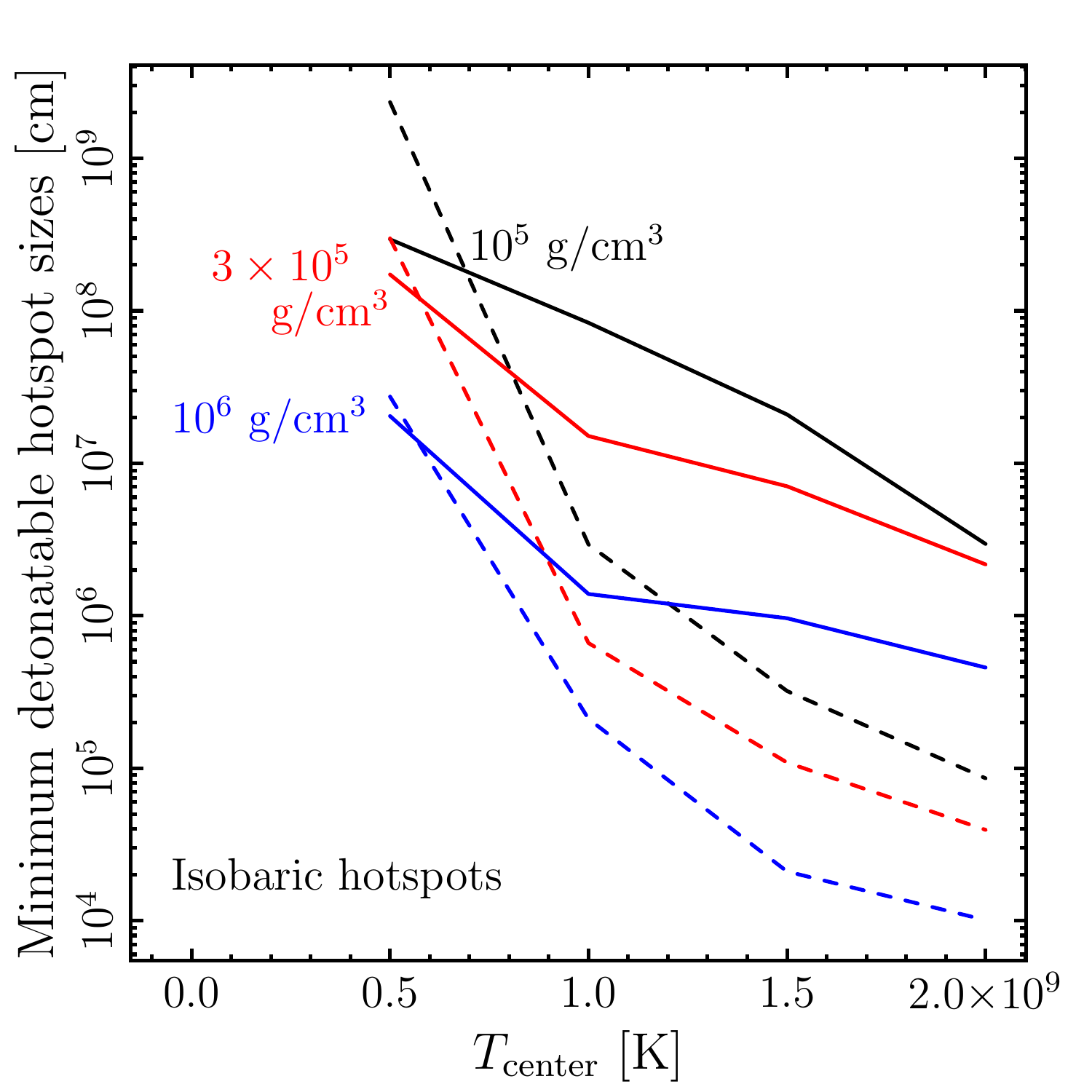}
	\caption{Same as Fig. \ref{fig:lcrit_isochoric}, but for isobaric hotspots.  There are no \cite{holc13a} data points for comparison, as their study did not consider isobaric hotspots.}
	\label{fig:lcrit_isobaric}
\end{figure}

In Figure \ref{fig:lcrit_isobaric}, we show results for constant pressure hotspots using the same labeling scheme as in Figure \ref{fig:lcrit_isochoric}.  As previously described, the increased ambient density yields a smaller ZND lengthscale, which in turn leads to much smaller detonatable hotspot sizes.  The pure He - aprox13 isobaric results approach the polluted - 206-isotope isochoric results, and the polluted - 206-isotope isobaric results are as much as $3000$ times smaller than the pure He - aprox13 isochoric results.  Our fiducial calculation at $T_{\rm center}=10^9$ K and $\rho_{\rm center} = 10^5 \cgsd$ now results in a detonation for a $3\E{6}$ cm hotspot, which is only 1\% the WD's scale height.  This bodes extremely well for the possibility of igniting He detonations during shell convection or during WD mergers.


\section{Lateral propagation of the detonation}
\label{sec:shellpropagation}

Now we investigate the steady-state structure of the laterally propagating surface detonation. Standard one-dimensional calculations of the structure of steady-state detonations need to be modified to account for the finite-gravity environment of the He envelope. Multidimensional simulations \citep{sim12,tmb12,moll13a} find steady He detonation velocities lower than the CJ velocity expected from fully-burned one-dimensional calculations, mainly due to the quenching effects of the curvature of the detonation front as well as post-shock radial expansion.

\citet{moor13a} investigated one-dimensional models of these effects, finding that there was a minimum He envelope mass that would support a steady detonation. Detonations in envelopes that were too small would experience significant quenching before a post-shock sonic locus was reached, preventing a self-sustaining detonation from forming. For envelopes large enough to support laterally propagating detonations, there is a one-to-one mapping between ambient conditions of an envelope (density, temperature, scale height, and composition) and a steady-state detonation solution. This allows us to identify an envelope structure with a single detonation length scale, $l_{\rm ZND}$ (here defined to be the length scale to $95$\% of the total energy release), and final-state nucleosynthesis. This final-state nucleosynthesis shows a strong dependence on envelope mass, with less massive envelopes producing mostly intermediate-mass elements and significant $^{56}$Ni production only for high-mass envelopes where the quenching effects act on time scales long enough to allow for significant burning to occur. In this section, we present calculations with an improved version of the code from \cite{moor13a}, using the 206-isotope nuclear reaction network described in the previous sections. 

\begin{figure}
        \plotone{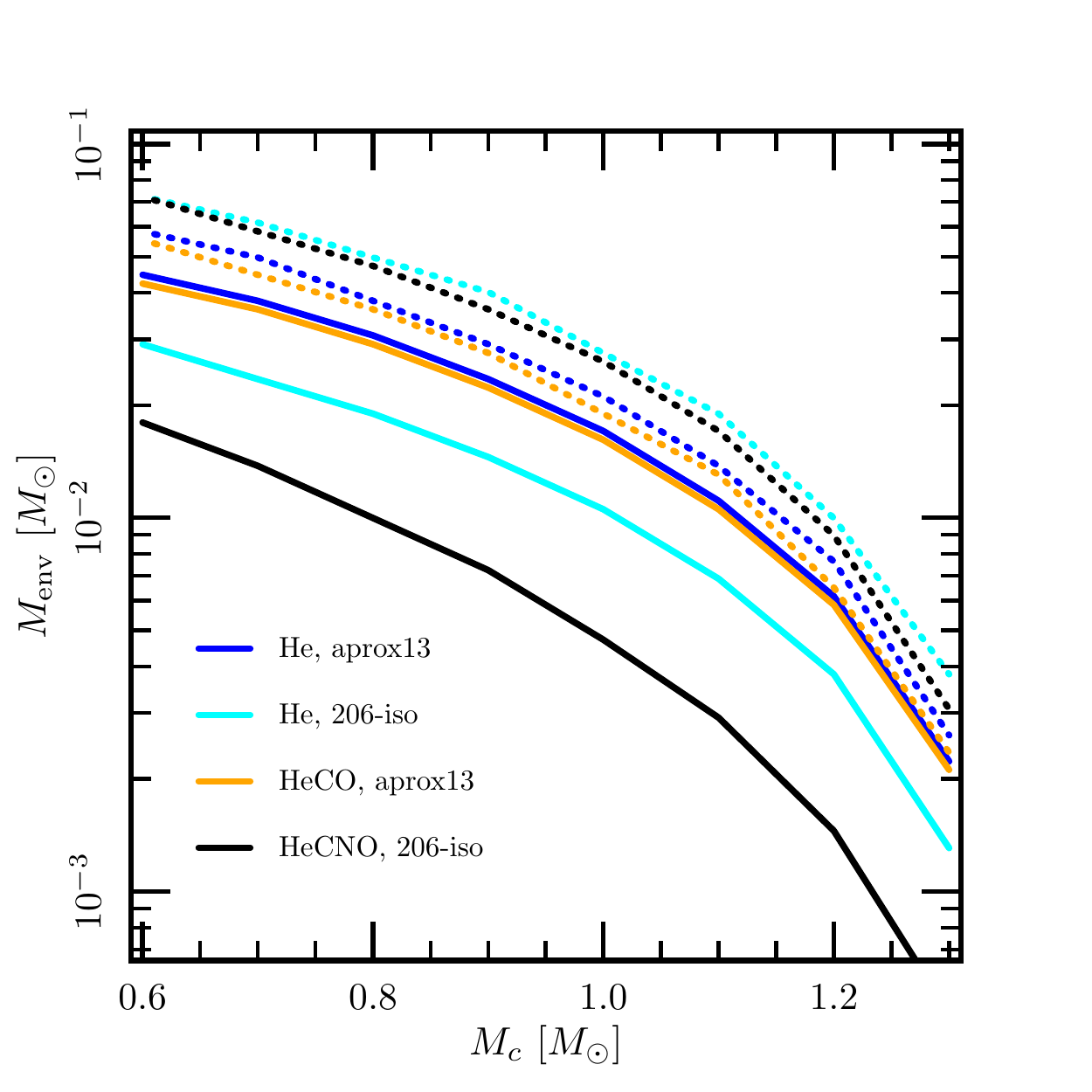}
        \caption{Minimum envelope masses, $M_{\rm env}$, that can support a steady detonation as a function of WD core mass, $M_c$, are shown in solid lines for several reaction network and composition combinations.  From top to bottom, the solid lines represent models with initial compositions of pure He and the aprox13 nuclear network (blue), pure He and a 206-isotope network (orange), a polluted composition of $X_{\rm 4He}=0.9$ and $X_{\rm 12C}=X_{\rm 16O}=0.05$ with the aprox13 network (cyan), and a composition of $X_{\rm 4He}=0.891$, $X_{\rm 12C}=X_{\rm 16O}=0.05$, and $X_{\rm 14N}=0.009$ with 206 isotopes (black).  The minimum shell masses where a significant portion of the ashes produced are radioactive on relevant timescales ($X_{\rm 48Cr} + X_{\rm 52Fe} + X_{\rm 56Ni} > 0.2$) are shown as dotted lines for the same reaction networks and compositions.  From top to bottom, the lines represent calculations with initially pure He and 206 isotopes (cyan), polluted He and 206 isotopes (black), pure He and aprox13 (blue), and polluted He and aprox13 (orange).}
        \label{fig:min_env_mass}
\end{figure}

Figure \ref{fig:min_env_mass} shows how the minimum envelope mass that supports a steady detonation depends on the WD core mass for different compositions and reaction networks. We see that using the larger 206-isotope reaction network with a pure He composition noticeably decreases the minimum detonatable envelope mass from the value predicted with the aprox13 network. Additionally, adding a more realistic fiducial envelope composition of $X_{\rm 4He}=0.891$, $X_{\rm 12C}=X_{\rm 16O} = 0.05$, and $X_{\rm 14N}=0.009$ reduces the detonatable envelope mass even more when used with the 206-isotope reaction network. However, similar polluted compositions ($X_{\rm 4He}=0.9$, $X_{\rm 12C}=X_{\rm 16O} = 0.05$) do not make a significant difference when used with aprox13, showing that the proton-catalyzed reactions near the beginning of the $\alpha$-chain have a significant effect on the overall steady-state detonation structure through the minimum detonatable envelope mass as well as the ignition length scales. It is therefore easier for such detonations to initiate as well as propagate when calculations employ more realistic reaction networks.

We also indicate where different calculations begin to produce significant amounts of isotopes that are radioactive on relevant timescales. The simple $\alpha$-chain reaction network produces radioactive $\alpha$-chain isotopes in slightly lower-mass envelopes than the more realistic 206-isotope network. Additionally, varying the initial composition in the 206-isotope networks has little effect on the amount of radioactive isotopes produced in larger envelopes, despite allowing for successful detonations in smaller envelopes. The larger networks and more realistic compositions also increase the range of $M_{\rm env}$ where we expect little radioactivity in the burning products.

The use of the realistic composition and network allows for successful detonations in He-rich shells with masses $< 10^{-2} \msol$.  This opens a new channel for double detonation SNe Ia in double C/O WD binaries, as C/O WDs are surrounded by thin He layers \citep{it85}.  While the accretors in these systems likely have very low mass He envelopes $\sim 10^{-3} \msol$, donors $\simeq 0.6 \msol$ possess He layers $\sim 10^{-2} \msol$, which may be enough to trigger a He detonation during the merging process.  This possibility was first suggested by \cite{rask12} and \cite{pakm13a}, although neither global study could resolve the initiation and propagation of the detonation in the He layer surrounding the accreting WD.  Future work on double detonation SNe Ia should allow for this channel, especially since there will be substantially more C/O pollution in the accreted He layer to further catalyze the detonation.

\begin{figure}
        \plotone{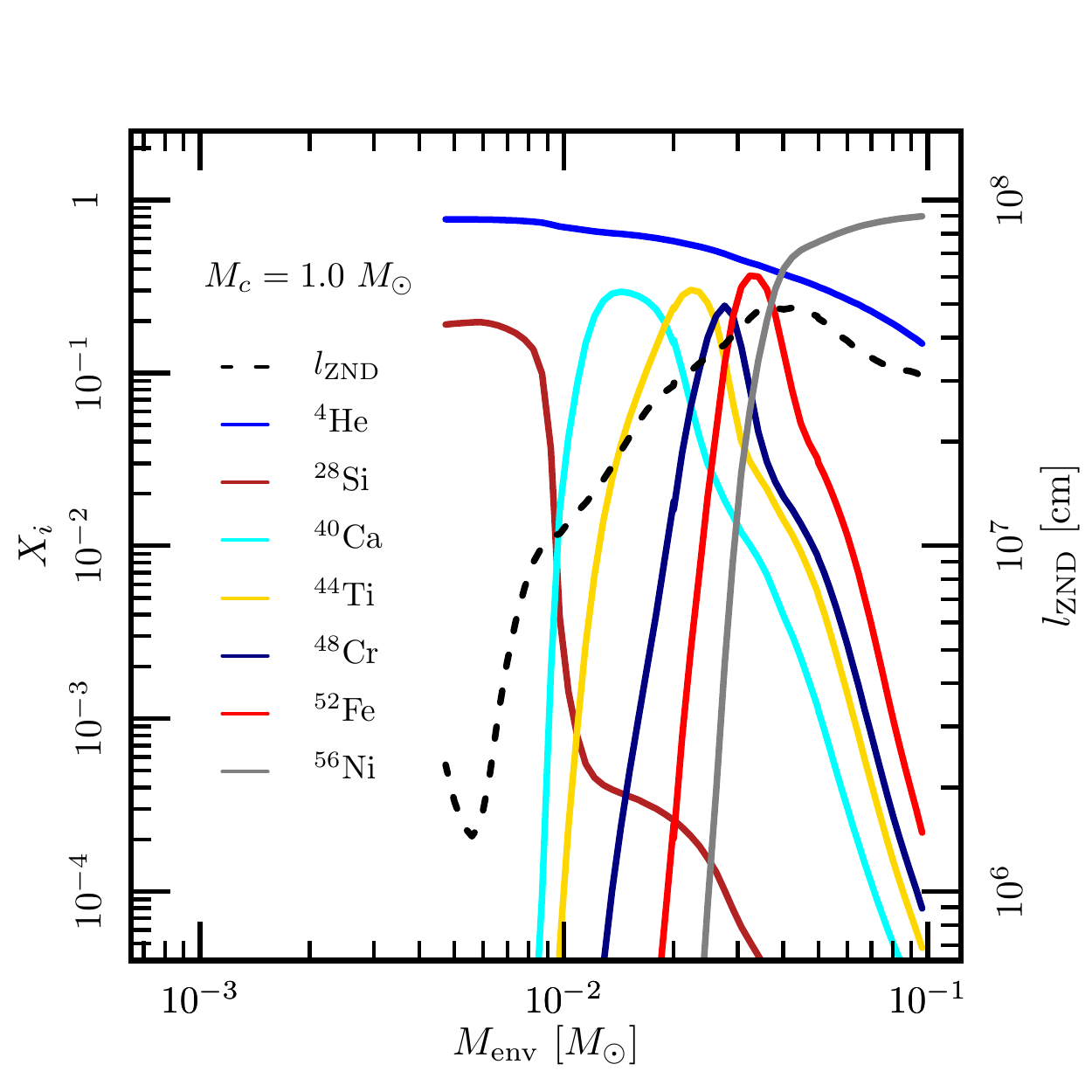}
        \caption{ZND length scales (dashed line - right axis) and final mass fractions (solid lines - left axis) of material burned by a steady, laterally-propagating detonation wave as a function of He envelope mass, $M_{\rm env}$, on a $1.0\ M_\odot$ C/O WD. Each envelope mass corresponds to a unique detonation velocity eigenvalue and final-state composition.  He is the most abundant element for the minimum-mass envelopes, followed by $^{28}$Si.  The remaining solid lines, from left to right in order of their first appearance above $X_i=10^{-4}$, represent the mass fractions of $^{40}$Ca, $^{44}$Ti, $^{48}$Cr $^{52}$Fe, and $^{56}$Ni.  Non-$\alpha$-chain isotopes are not shown.}
        \label{fig:final_abundances}
\end{figure}

To get a picture of how the specific abundances are evolving as we vary the envelope mass, Figure \ref{fig:final_abundances} shows the sensitive dependence of the detonation products on the mass of the envelope for our fiducial polluted envelope composition on a $1.0 \msol$ WD core. Detonations in the lowest-mass envelopes are dominated by $^{28}$Si, $^{40}$Ca, and unburned He. Significant radioactivity requires higher-mass envelopes, with $^{56}$Ni only being produced by detonations in the most massive of envelopes. We define the detonation length scale, $l_{\rm ZND}$, as the post-shock distance to $95$\% of the total energy release. Since these are all pathological detonations, this typically occurs in the section of the post-shock flow that is supersonic relative to the detonation front, so not all of the energy produced can propel the detonation. As the envelope mass increases, the detonations get closer to the limit of a Chapman-Jouget detonation, where all the burning occurs in material that is sonically connected to the detonation front. The qualitative behavior of such nucleosynthesis is the same for other WD core masses, with the minimum detonatable mass shifted depending on $M_c$.

The result that the minimum-mass detonatable shells produce intermediate-mass elements while avoiding iron-group element production is particularly interesting for double detonation SNe Ia from double WD binaries, as the He detonation will likely be triggered in the minimum-mass He shell.  In merging systems, the ease of initiating the detonation may mean that the barrier to a successful detonation lies in the size of the shell, so that a fully propagating detonation is realized once the minimum He mass has been transferred from the donor.  In stable mass transfer systems with a He donor, the accretion rate decreases from initially high values $\dot{M} \sim 10^{-6} \msol$ yr$^{-1}$, and the system evolves from burning He stably to unstably.  As $\dot{M}$ decreases, the unstable He-burning events become more and more violent, until the size of the shell is large enough to support a propagating detonation.  Thus, the first detonation in this stable mass transfer channel will also occur in the smallest detonatable shell.

If this minimum shell He detonation triggers a core detonation and a subsequent SN Ia, the He shell ashes will not yield large amounts of high-velocity iron-group elements, which have been ruled out by several studies (e.g., \citealt{nuge97,krom10}).  The minimum-mass shells will instead consist of high-velocity unburnt He, $^{28}$Si, $^{40}$Ca, and a small amount of unburnt primordial metals.   Given the lack of non-thermal electrons, the He will likely stay neutral and unobservable.  The remaining elements may explain observations of high-velocity absorption features seen in most SNe Ia \citep{mazz05,tana08a,blon12a,chil14a,magu14a}.  Further work is necessary to determine if the strong abundance enhancements of the $^{28}$Si, $^{40}$Ca, and other metals are enough to produce these features.  However, given that they comprise essentially all of the opaque material at these velocities, this is a very promising avenue of future research.


\section{Conclusions}

In this paper, we have shown that the ignition of He detonations and their propagation in WD envelopes are significantly impacted by the inclusion of a small amount of C/O pollution and a full nuclear reaction network.  Motivated by an examination of the relevant reaction rates in Section \ref{sec:motivation} and hotspot conditions in Section \ref{sec:expectations}, our calculations in Sections \ref{sec:hotspotsizecalculation} and \ref{sec:shellpropagation} indicate that He detonation ignition and propagation is possible during double WD mergers and in convectively burning envelopes on WD surfaces and is thus a promising channel for SNe ``.Ia'' and Ia.

Detonations in low-mass He envelopes may produce high-velocity Si, Ca, and unburned metal spectral features if a double detonation occurs and the core explodes as a SN Ia.  If the He envelope detonates without triggering a secondary detonation in the WD core (either due to weak shock focusing or a core composition of O/Ne), then it could appear as a ``.Ia'' supernova if significant radioactive isotopes are produced. Otherwise, the low ejecta mass coupled with little to no radioactivity in the ashes would produce a virtually unobservable explosion, perhaps detectable as circumstellar material if illuminated by subsequent explosions.

Given the relative infancy of work regarding He detonation ignition and propagation, it is clear that further study is required.  Future work will include multidimensional hydrodynamic calculations of both the initiation and propagation of the detonation, a more rigorous description of the initial hotspot parameters, quantification of the production of high-velocity spectral features expected from the burned shell, and an exploration of a range of spatially varying initial compositions.  If the compositional gradient between the He-rich envelope and the C/O-rich core is gradual enough, the He-burning detonation may transition directly to a C-burning flame in an edge-lit double detonation SN Ia scenario, without the need for a convergence-driven core ignition.  An edge-lit core ignition is also possible if the He detonation reaches full strength well above the core-envelope interface.  The small detonatable hotspots we have found now allow for this possibility, as they do not require a large fraction of the envelope's scale height to form.


\acknowledgments

We thank Lars Bildsten, James Guillochon, Falk Herwig, Bill Paxton, Eliot Quataert, Ivo Seitenzahl, Frank Timmes, and Dean Townsley for helpful discussions and the referee for their comments.  KJS acknowledges the hospitality of Caskhouse, where some of these calculations were performed.  KJS is supported by NASA through Einstein Postdoctoral Fellowship grant number PF1-120088 awarded by the Chandra X-ray Center, which is operated by the Smithsonian Astrophysical Observatory for NASA under contract NAS8-03060.  KM is partially supported by the National Science Foundation under grant AST 11-09174.



\end{document}